# Compressed Air Energy Storage-Part II: Application to Power System Unit Commitment

Junpeng Zhan, *Member, IEEE,* Yunfeng Wen, *Member, IEEE,*
Osama Aslam Ansari, *Student Member, IEEE,* and C. Y. Chung, *Fellow, IEEE*

***Abstract*—Unit commitment (UC) is one of the most important power system operation problems. To integrate higher penetration of wind power into power systems, more compressed air energy storage (CAES) plants are being built. Existing cavern models for the CAES used in power system optimization problems are not accurate, which may lead to infeasible solutions, e.g., the air pressure in the cavern is outside its operating range. In this regard, an accurate CAES model is proposed for the UC problem based on the accurate bi-linear cavern model proposed in the first paper of this two-part series. The minimum switch time between the charging and discharging processes of CAES is considered. The whole model, i.e., the UC model with an accurate CAES model, is a large-scale mixed integer bi-linear programming problem. To reduce the complexity of the whole model, three strategies are proposed to reduce the number of bi-linear terms without sacrificing accuracy. McCormick relaxation and piecewise linearization are then used to linearize the whole model. To decrease the solution time, a method to obtain an initial solution of the linearized model is proposed. A modified RTS-79 system is used to verify the effectiveness of the whole model and the solution methodology.**

## NOMENCLATURE

### Sets/Indices

|  |  |
| --- | --- |
| $b$, $\Omega_B$ | Bus index and set of all bus indices, respectively |
| $i$ | Index for injections (including conventional generation units, wind generation units, compressed air energy storage (CAES)) |
| $j$, $\Omega_J$ | Scenario index and set of all scenario indices, respectively |
| $l$, $\Omega_L$ | Line index and set of all line indices, respectively |
| $t$, $\Omega_T$ | Time index and set of all time indices, respectively |
| $\Omega_{T0}$ | $\{0,1,2,\cdots,n_t-1\}$ where $n_t$ represents the number of time periods |
| $\Omega_{T1}$ | $\{1,2,\cdots,n_t\}$ |
| $I_b$ | Set of indices of injections connected to bus $b$ |
| $I_c$ | Set of conventional generation units |
| $I_d$ | Set of all load indices |
| $I_s$ | Set of CAES units |
| $I_w$ | Set of wind generation units |
| $J^t$ | Set of indices of all scenarios considered at time $t$ |
| $I^{tj_2}$ | Indices of all units available for dispatch in scenario $j_2$ at time $t$ |

### Parameters

|  |  |
| --- | --- |
| $c_v$ | Constant volume specific heat (J/(kg K)) |
| $d^{tij}$ | Load demand (MW) |
| $f_l^{max}$ | The maximum power flow of line $l$ (MW) |
| $h_c$ | Heat transfer coefficient (W/(m$^2$ K)) |
| $k$ | A constant equal to 1.4 |
| $m_{av0}$ | Average mass of air in the cavern (kg) |
| $p_{in}$ | Pressure of the air charged into a cavern (bar) |
| $p_{\max}^i$, $p_{\min}^i$ | Maximum and minimum pressures in a cavern for optimal operation of CAES (bar) |
| $A_c$ | Surface area of the cavern wall (m$^2$) |
| $C_v^{ti}$, $C_w^{ti}$ | Startup and shutdown costs, respectively ($) |
| $C_{\delta+}^{ti}$, $C_{\delta-}^{ti}$ | Cost coefficients of upward and downward load-following ramp reserve, respectively ($/MW) |
| $C_{Pa}^{ti}$, $C_{Pb}^{ti}$ | Cost coefficients of conventional generators |
| $C_{sc}^{ti}$, $C_{sd}^{ti}$ | Charging and discharging costs, respectively ($/MWh) |
| $C_{ws}^{ti}$ | Wind shedding cost ($/MWh) |
| $C_R^t$ | Reserve cost ($/MWh) |
| $M_{i,l}$ | The element in the $i$th row and the $l$th column of a node-branch incidence matrix |
| $P_{ch}^{i,\max}$, $P_{ch}^{i,\min}$ | Maximum and minimum charging power of CAES, respectively (MW) |
| $P_{dch}^{i,\max}$, $P_{dch}^{i,\min}$ | Maximum and minimum discharging power of CAES, respectively (MW) |
| $P_R^t$ | Power reserve required (MW) |
| $R$ | Gas constant (bar $\cdot$ m$^3$ $\cdot$ kg$^{-1}$ $\cdot$ K$^{-1}$) |
| $T_{in}$ | Temperature of the air injected into a cavern (K) |
| $T_{RW}$ | Temperature of the cavern wall (K) |
| $T_{\max}^i$, $T_{\min}^i$ | Maximum and minimum temperature of air inside a cavern (K) |
| $V_s$ | Volume of the storage (m$^3$) |
| $W_{\max}^{tij}$ | Maximum wind power that can be generated at scenario $j$ (MW) |
| $\tilde{\gamma}_l$ | Susceptance of a line on right-of-way $l$ (Siemens) |
| $\delta_{\max+}^{ti}$, $\delta_{\max-}^{ti}$ | Upward and downward ramping limits, respectively (MW) |
| $\psi^{tj}$ | Probability of scenario $j$ at time $t$ |
| $\Delta t$ | Time interval (second) |

### Variables

|  |  |
| --- | --- |
| $d_{ls}^{tij}$ | Load shedding (MW) |
| $f_l$ | Total active power flow on line $l$ (MW) |
| $\dot{m}_{in}^{tij}$, $\dot{m}_{out}^{tij}$ | Rate of flow of air mass charged into and discharged from a cavern, respectively (kg/s) |
| $p_s^{tij}$, $T_s^{tij}$, $m_s^{tij}$ | Pressure (bar), temperature (K), and mass (kg) of air stored in the cavern, respectively |

The work was supported in part by the Natural Sciences and Engineering Research Council (NSERC) of Canada and the Saskatchewan Power Corporation (SaskPower).

The authors were with the Department of Electrical and Computer Engineering, University of Saskatchewan, Saskatoon, SK S7N 5A9, Canada (e-mail: zhanjunpeng@gmail.com, y.f.wen@usask.ca, oa.ansari@usask.ca, c.y.chung@usask.ca).

$p_{s,(\mathrm{xx})}^{(t+1)ij}$ Pressure after a (xx) process where (xx) can be 'ch', 'dch', and 'idl', which represent charging, discharging, and idle, respectively (bar)

$u^{ti}$ Unit on/off status; 1 if unit is on, 0 otherwise

$v^{ti}$, $w^{ti}$ Binary startup and shutdown states, respectively; 1 if unit $i$ has startup/shutdown events, 0 otherwise

$P^{tij}$ Power output from conventional unit $i$ (MW)

$P_{\mathrm{ch}}^{tij}$, $P_{\mathrm{dch}}^{tij}$ Charging and discharging power, respectively (MW)

$T_{s,(\mathrm{xx})}^{(t+1)ij}$ Temperature after a (xx) process where (xx) can be 'ch', 'dch', and 'idl', which represent charging, discharging, and idle, respectively (K)

$T_{s,\mathrm{chdch}}^{(t+1)ij}$ Temperature at time $(t+1)$ if the $t$th period is either a charging or discharging process (K)

$W^{tij}$ Scheduled wind power generation (MW)

$\alpha_{c,i}^{\kappa,t}$, $\beta_{c,i}^{\kappa,t}$ Binary variable indicating the charging and discharging processes, respectively

$\delta_{+}^{ti}$, $\delta_{-}^{ti}$ Upward and downward load-following ramping reserve needed from unit $i$ at time $t$ for transition to time $t+1$, respectively (MW)

$\theta_{l,\mathrm{fr}}^{\kappa}$ Phase angle of from-side node of right-of-way $l$ (rad)

$\theta_{l,\mathrm{to}}^{\kappa}$ Phase angle of to-side node of right-of-way $l$ (rad)

## I. INTRODUCTION

UNIT commitment (UC) is a key power system operation problem [1]-[4] that determines the unit on/off status ahead of time to supply sufficient electric power to customers in a secure and economic manner. A comprehensive review of UC is provided in [5] and [6].

It is beneficial to integrate energy storage systems into UC problems [3], [7]-[9]. To hedge the wind power output uncertainty, pumped-storage units are incorporated in the UC problem [7]. Reference [8] proposed deterministic and interval UC formulations for the co-optimization of controllable generation and pumped hydro energy storage. In [3] and [9], fast-response battery energy storage is utilized in UC problems for congestion relief and frequency support, respectively.

Compressed air energy storage (CAES), as mentioned in the first paper of this two-part series, is a promising large-scale energy storage technology. CAES has been used to enhance power system operation by mitigating wind shedding [10], smoothing wind power fluctuation [11], providing ancillary service [12], participating in energy and reserve markets [13], [14], etc.

Some preliminary work considering CAES in UC problems has been done [15]-[18]. Reference [15] integrates ideal and generic storage devices into stochastic real-time UC problems to deal with the stochasticity and intermittence of non-dispatchable renewable resources. Reference [16] developed an enhanced security constrained UC formulation considering CAES and wind power. In [17], CAES and sodium sulfur batteries are used in a UC problem to maximize the wind energy penetration level. In [18], a constant-pressure CAES is modeled for the bi-level planning of a microgrid including CAES, where UC with CAES is described on the lower level.

In the papers mentioned above, the temperature of the air in the cavern of CAES is assumed to be constant (called constant-temperature cavern model for the CAES). The pressure of the air is then a linear function of the mass of air in the cavern according to the ideal gas law. The pressure of the air in a CAES cavern must be within an operating range to ensure stable CAES operation. However, solutions obtained from the constant-temperature cavern model can allow the pressure of the air in the cavern to fall outside of the operating range. That is, the constant-temperature cavern model is inaccurate and may result in an infeasible solution.

As mentioned in the first paper of this two-part series, accurate analytical models [19] that have been proposed for the cavern used in CAES are highly non-linear and therefore cannot be integrated into large-scale power system optimization problems. In this regard, the bi-linear accurate cavern model proposed in the first paper of this series is integrated into power system operation problems in this second paper to ensure the pressure of the air in the cavern is maintained within the operating range. This is an important and urgent task considering two CAES plants are already in operation and several more plants are under construction, as mentioned in the first paper. This second paper focuses on integrating the CAES into UC problems. However, the CAES model proposed herein can be easily extended to other power system optimization problems, e.g., optimal power flow, economic dispatch, etc.

In CAES, a single motor/generator set is used to drive both the compressor and expander. Therefore, it needs time to switch between the charging and discharging processes. In the literature, constraints associated with the minimum switch time between charging and discharging processes are usually not considered. Reference [17] proposed a set of constraints to ensure switch time. In the current paper, a novel method with a smaller number of constraints and variables than [17] is proposed to ensure the minimum switch time.

In the proposed CAES model using the accurate bi-linear cavern model, there are two kinds of bi-linear terms, i.e., the product of a binary variable and a continuous variable (called a binary-continuous bi-linear term) and the product of two continuous variables (called a continuous bi-linear term). These bi-linear terms complicate the whole model, i.e., the UC model with CAES using the accurate bi-linear cavern model.

To decrease the complexity of solving the whole model, three strategies are proposed to reduce the number of binary-continuous and continuous bi-linear terms. Given that piecewise linearization of continuous bi-linear terms will introduce more constraints and binary variables, these strategies can significantly reduce the number of constraints and binary variables added to the original whole model. Therefore, these strategies can significantly reduce the solution time.

This paper uses a McCormick relaxation [9] to replace the binary-continuous bi-linear term by a new continuous variable subject to several linear constraints. The advantages of the McCormick relaxation for binary-continuous bi-linear terms include no error, i.e., it is a tight relaxation, and no introduction of new binary/integer variables to the original model.

In the CAES model, the mass, pressure, and temperature of the air in the cavern are involved in the continuous bi-linear terms and have large ranges. Unfortunately, the McCormick relaxation for continuous bi-linear terms has a relatively large error when the ranges of continuous variables are large. Therefore, the McCormick relaxation is not applicable to linearize the continuous bi-linear terms herein. Piecewise linearization is a widely used and effective method to approximate a non-linear function [20]. Therefore, the piecewise linearization is used to linearize the continuous bi-linear terms. Specifically, the continuous bi-linear term is transformed into the difference of two quadratic terms, which are subsequently piecewise linearized.

It is always valuable to supply a mixed integer linear programming (MILP) solver with an initial solution to solve an MILP problem, which can reduce the solution time and increase the solution accuracy [21]. In this regard, a method to obtain an initial solution for the linearized whole model is proposed that utilizes the solution obtained from the UC with CAES using a constant-temperature cavern model; it can significantly reduce the solution time required to solve the linearized whole model.

In summary, the contributions of this second paper include a novel bi-linear CAES model for the UC and its linearization, strategies to simplify the bi-linear CAES model for the UC, and a method to obtain an initial solution for the linearized whole model.

The rest of the paper is organized as follows. Section II details the UC model considering CAES. Section III describes the model reformulation and solution method used to solve the whole UC model. Simulation results are given in Section IV and conclusions are drawn in Section V.

## II. Unit Commitment Considering Compressed Air Energy Storage

### A. Unit Commitment Model

In this subsection, the UC model proposed in [22] is adopted and modified to include CAES.

#### 1) Objective Function

The objective function is given in (1). The first term is the startup and shutdown costs of conventional generators; the second term is the cost of load-following ramp reserves of conventional generators; the third term includes the power generation cost, charging and discharging costs, and the penalty cost of wind shedding; and the last term is the cost of spinning reserves.

$$f_{\text{obj}} = \sum_{t\in\Omega_T}\sum_{i\in I_{\text{c}}}(C_v^{ti}v^{ti} + C_w^{ti}w^{ti}) + \sum_{t\in\Omega_T}\sum_{i\in I_{\text{c}}}[C_{\delta+}^{ti}(\delta_+^{ti}) + C_{\delta-}^{ti}(\delta_-^{ti})] + \sum_{t\in\Omega_T}\sum_{j\in J^t}\psi^{tj}\big[\sum_{i\in I_{\text{c}}}\big(C_{Pa}^{ti}P^{tij} + C_{Pb}^{ti}\big) + \sum_{i\in I_{\text{s}}}(C_{\text{sc}}^{ti}P_{\text{ch}}^{tij} + C_{\text{sd}}^{ti}P_{\text{dch}}^{tij}) + \sum_{i\in I_{\text{w}}}C_{\text{ws}}^{ti}\big(W_{\max}^{tij} - W^{tij}\big)\big] + \sum_{t\in\Omega_T}\sum_{j\in J^t}\psi^{tj}\big[\sum_{i\in I_{\text{c}}}C_{\text{R}}^{t}(u^{ti}P_{\max}^{ti} - P^{tij}) + \sum_{i\in I_{\text{s}}}C_{\text{R}}^{t}\big(\beta^{tij}P_{\text{dch}}^{i,\max} - P_{\text{dch}}^{tij}\big)\big] \quad (1)$$

#### 2) DC Power Flow

Constraint (2) represents the power balance at each bus and (3) represents the DC power flow [22]. Constraint (4) represents the capacity limit of each transmission line.

$$\sum_{l\in\Omega}M_{b,l}\,f_l^{tj} + \sum_{i\in b}\big(P^{tij} + W^{tij} - d^{tij} + d_{\text{ls}}^{tij} + P_{\text{dch}}^{tij} - P_{\text{ch}}^{tij}\big) = 0,\ \ \forall b\in\Omega_B, \forall t\in\Omega_T \quad (2)$$

$$f_l^{tj} - \tilde{\gamma}_l\big(\theta_{l,\text{fr}}^{tj} - \theta_{l,\text{to}}^{tj}\big) = 0, \forall l\in\Omega_L, \forall t\in\Omega_T \quad (3)$$

$$\big|f_l^{tj}\big| \le f_l^{\max},\ \forall l\in\Omega_L\ ,\ \forall t\in\Omega_T \quad (4)$$

#### 3) Spinning Reserve

The spinning reserve can be expressed as (5) [23]. Note that $P_{\text{ch}}^{i}$ is a load but it does not appear on the right-hand side of (5) because the charging load of CAES can be curtailed immediately when reserve is required.

$$\sum_{i\in I_{\text{c}}}u^{ti}P_{\max}^{ti} + \sum_{i\in I_{\text{w}}}W^{tij} + \sum_{i\in I_{\text{s}}}\beta^{tij}P_{\text{dch}}^{i,\max} \ge \sum_{i\in I_{\text{d}}}(d^{tij}) + P_R^t, \forall t\in\Omega_T, \forall j\in\Omega_J \quad (5)$$

where $P_R^t$ represents the required spinning reserve at time $t$, which is set to the power capacity of the largest unit in the system.

#### 4) Intertemporal Constraints:

Load-following ramping limits and reserves are given in (6)-(9) according to [22].

$$0 \le \delta_+^{ti} \le \delta_{\max+}^{ti} \quad (6)$$

$$0 \le \delta_-^{ti} \le \delta_{\max-}^{ti} \quad (7)$$

$$P^{tij_2} - P^{(t-1)ij_1} \le \delta_+^{(t-1)i}, j_1\in J^{t-1}, j_2\in J^t \quad (8)$$

$$P^{(t-1)ij_1} - P^{tij_2} \le \delta_-^{(t-1)i}, j_1\in J^{t-1}, j_2\in J^t \quad (9)$$

The other constraints used include the startup and shutdown constraints, the minimum up and down times for conventional generators [22], and the lower and upper bounds for the output of conventional generators and wind farms.

### B. CAES Constraints

The mass flow rate in and out, i.e., $\dot{m}_{\text{in}}^{tij}$ and $\dot{m}_{\text{out}}^{tij}$, can be expressed as linear functions of the charging power ($P_{\text{ch}}^{tij}$) and discharging power ($P_{\text{dch}}^{tij}$), respectively, according to [24]:

$$\dot{m}_{\text{in}}^{tij} = c_{\text{Ain}}P_{\text{ch}}^{tij},\ \ \ \forall t\in\Omega_{T1}, i\in\Omega_s, \forall j \quad (10)$$

$$\dot{m}_{\text{out}}^{tij} = c_{\text{Aout}}P_{\text{dch}}^{tij},\ \ \forall t\in\Omega_{T1}, i\in\Omega_C, \forall j, \quad (11)$$

where the values of the coefficients $c_{\text{Ain}}$ and $c_{\text{Aout}}$ are adopted from [24].

There is an optimal operating range for the pressure of the air in the cavern, which can be expressed as

$$p_{\min}^{i} \le p_s^{tij} \le p_{\max}^{i},\ \ \forall t\in\Omega_{T1}\ . \quad (12)$$

The CAES cannot be in charging and discharging processes at the same time, which can be modeled as

$$\alpha^{tij} + \beta^{tij} \le 1,\ \ \ \forall t\in\Omega_{T1}, \forall i\in\Omega_C, \forall j \quad (13)$$

where $\alpha^{tij}$ and $\beta^{tij}$ are binary variables used to represent the charging and discharging processes, respectively. The idle process can be represented as $(1-\alpha^{tij}-\beta^{tij})$ as the CAES should be in one and only one of the charging, discharging, and idle processes at a time. This representation can reduce the number of variables and equality constraints compared to using another binary variable to indicate the status of the idle process.

The lower and upper bounds of the charging power and discharging power can be expressed as (14) and (15), respectively. If the CAES is not in the charging (discharging) process, then $\alpha^{tij}=0$ ($\beta^{tij}=0$) and therefore the charging (discharging) power is 0.

$$\alpha^{tij}P_{\text{ch}}^{i,\min} \le P_{\text{ch}}^{tij} \le \alpha^{tij}P_{\text{ch}}^{i,\max},\ \forall t\in\Omega_{T1}, \forall i\in\Omega_C \quad (14)$$

$$\beta^{tij}P_{\text{dch}}^{i,\min} \le P_{\text{dch}}^{tij} \le \beta^{tij}P_{\text{dch}}^{i,\max},\ \forall t\in\Omega_{T1}, \forall i\in\Omega_C \quad (15)$$

The following constraint ensures that, when it is in idle process, i.e., $P_{\text{ch}}^{tij}=0$ and $P_{\text{dch}}^{tij}=0$, the indicator for the idle process is equal to 1, i.e., $1-\alpha^{tij}-\beta^{tij}=1$. Note that $P_{\text{ch}}^{i,\min}>0$ and $P_{\text{dch}}^{i,\min}>0$, i.e., $\min\big(P_{\text{ch}}^{i,\min}, P_{\text{dch}}^{i,\min}\big)>0$.

$$(\alpha^{tij}+\beta^{tij})\cdot\min\big(P_{\text{ch}}^{i,\min}, P_{\text{dch}}^{i,\min}\big) \le P_{\text{ch}}^{tij} + P_{\text{dch}}^{tij}, \forall t\in\Omega_{T1} \quad (16)$$

### C. Minimum Switch Time Between Charging and Discharging

In a CAES plant, a single motor/generator set is used to drive both the compressor and expander. Thus, a minimum time, i.e., 20 minutes, is required to switch between charging and discharging [24]. The following two constraints are proposed to guarantee this switch time:

$$\alpha^{tij} + \beta^{(t+t_1)ij} \leq 1,\ t = 1,2,\cdots,n_t - a_1,\ t_1 = 1,\cdots,a_1 \tag{17}$$

$$\alpha^{(t+t_1)ij} + \beta^{tij} \leq 1,\ t = 1,2,\cdots,n_t - a_1,\ t_1 = 1,\cdots,a_1 \tag{18}$$

where $a_1$ is an integer and $a_1 = \frac{\text{20 minutes}}{\text{time interval}}$. Compared to [17], these switch time constraints do not introduce extra binary/integer variables and the number of constraints is reduced. That is, these constraints are simpler, which can reduce the complexity of the CAES model.

*D. Temperature and Pressure Models During Charging, Discharging, and Idle Processes*

When $t = 0$, $m_s^{tij}$ ($T_s^{tij}$, $p_s^{tij}$) represents the initial mass (temperature, pressure) of the air in the cavern and is the same for each scenario $j$. All of the other notations with superscript $t \neq 0$ or $(t+1) \neq 0$ in (19)-(28) and (43)-(47) are variables. For the mass ($m_s^{tij}$), temperature ($T_s^{tij}$), and pressure ($p_s^{tij}$) of air, the values are instantaneous. For all of the other variables involved in (19)-(28) and (43)-(47), the values are assumed to be constant for a given period of time.

According to the first paper of this two-part series, the temperature (pressure) of the air in the cavern in the charging, discharging, and idle processes can be expressed as (19), (21), and (23) ((20), (22), and (24)), respectively.

$$-m_s^{tij}T_{s,\text{ch}}^{(t+1)ij} + T_s^{tij}m_s^{tij} + c_1\dot{m}_{\text{in}}^{(t+1)ij}T_s^{tij} + c_2 m_s^{tij}\dot{m}_{\text{in}}^{(t+1)ij} + (\mathrm{c}_3)\dot{m}_{\text{in}}^{(t+1)ij} - \Delta t c_4 T_s^{tij} = -c_4 T_{RW}\Delta t,\quad \forall t \in \Omega_{T0} \tag{19}$$

$$-m_s^{tij}p_{s,\text{ch}}^{(t+1)ij} + p_s^{tij}m_s^{tij} + (k-1)\Delta t p_s^{tij}\dot{m}_{\text{in}}^{(t+1)ij} + a_2\Delta t k m_{av0}^{k-1} m_s^{tij}\dot{m}_{\text{in}}^{(t+1)ij} + (c_5)\dot{m}_{\text{in}}^{(t+1)ij}T_s^{tij} + (c_6)m_s^{tij} + (c_7)\dot{m}_{\text{in}}^{(t+1)ij} - (c_4)\Delta t p_s^{tij} = 0,\ \forall t \in \Omega_{T0} \tag{20}$$

$$-m_s^{tij}T_{s,\text{dch}}^{(t+1)ij} + m_s^{tij}T_s^{tij} + (c_8)T_s^{tij}\dot{m}_{\text{out}}^{(t+1)ij} - (c_4)\Delta t T_s^{tij} = -(c_4)\Delta t T_{RW},\ \forall t \in \Omega_{T0} \tag{21}$$

$$-m_s^{tij}p_{s,\text{dch}}^{(t+1)ij} + m_s^{tij}p_s^{tij} - k\Delta t\dot{m}_{\text{out}}^{(t+1)ij}p_s^{tij} - \frac{(c_4)R}{V_s^i}\Delta t m_s^{tij}T_s^{tij} + (c_9)T_s^{tij}\dot{m}_{\text{out}}^{(t+1)ij} + \frac{(c_4)R}{V_s^i}T_{RW}\Delta t m_s^{tij} - \frac{(c_4)R}{V_s^i}T_{RW}\Delta t^2 0.5\dot{m}_{\text{out}}^{(t+1)ij} = 0,\ \forall t \in \Omega_{T0} \tag{22}$$

$$\frac{a_4}{m_{av0}}e^{-a_4}m_s^{tij}T_s^{tij} - T_{s,\text{idl}}^{(t+1)ij} - T_{RW}\frac{a_4}{m_{av0}}e^{-a_4}m_s^{tij} + (c_{10})T_s^{tij} = T_{RW}(c_{10} - 1),\ \forall t \in \Omega_{T0} \tag{23}$$

$$\frac{a_4}{m_{av0}}e^{-a_4}m_s^{tij}p_s^{tij} - \frac{a_4}{m_{av0}}e^{-a_4}\frac{RT_{RW}}{V_s^i}m_s^{tij}m_s^{tij} - p_{s,\text{idl}}^{(t+1)ij} + (c_{10})p_s^{tij} + c_{11}m_s^{tij} = 0,\ \forall t \in \Omega_{T0} \tag{24}$$

where $a_2$-$a_4$ and $c_1$-$c_{11}$ are parameters defined in the Appendix.

The first 4, 5, 3, 5, 1, and 2 terms in (19), (20), (21), (22), (23), and (24), respectively, are bi-linear terms.

*E. Relationship Between Two Consecutive Time Periods for Temperature, Pressure, and Mass of Air in the Cavern*

The temperature and pressure of the air in the cavern at time $t+1$ can be expressed using (25) and (26), respectively, which are equal to the values in the charging, discharging, or idle processes according to the values of $\alpha_c^{(t+1)ij}$ and $\beta_c^{(t+1)ij}$.

$$T_s^{(t+1)ij} = \alpha^{(t+1)ij}T_{s,\text{ch}}^{(t+1)ij} + \beta^{(t+1)ij}T_{s,\text{dch}}^{(t+1)ij} + \left(1 - \alpha^{(t+1)ij} - \beta^{(t+1)ij}\right)T_{s,\text{idl}}^{(t+1)ij},\forall t \in \Omega_{T0} \tag{25}$$

$$p_s^{(t+1)ij} = \alpha^{(t+1)ij}p_{s,\text{ch}}^{(t+1)ij} + \beta^{(t+1)ij}p_{s,\text{dch}}^{(t+1)ij} + (1 - \alpha^{(t+1)ij} - \beta^{(t+1)ij})p_{s,\text{idl}}^{(t+1)ij},\forall t \in \Omega_{T0} \tag{26}$$

The relationship between the mass of air in the cavern at two consecutive time intervals can be expressed as

$$m_s^{(t+1)ij} = m_s^{tij} + \alpha^{(t+1)ij}\dot{m}_{\text{in}}^{(t+1)ij}\Delta t - \beta^{(t+1)ij}\dot{m}_{\text{out}}^{(t+1)ij}\Delta t,\ \forall t \in \Omega_{T0} \tag{27}$$

Note that it is guaranteed by (10), (11), and (13)-(15) that $\dot{m}_{\text{out}}^{(t+1)ij} = 0$ in the charging process and $\dot{m}_{\text{in}}^{(t+1)ij} = 0$ in the discharging process. Therefore, $\alpha^{(t+1)ij}$ and $\beta^{(t+1)ij}$ in (27) can be deleted, i.e., the bi-linear constraint (27) is equivalent to the linear constraint:

$$m_s^{(t+1)ij} = m_s^{tij} + \dot{m}_{\text{in}}^{(t+1)ij}\Delta t - \dot{m}_{\text{out}}^{(t+1)ij}\Delta t, t \in \Omega_{T0} \tag{28}$$

Therefore, the **optimization model** of UC considering CAES is complete and can be formed as

Minimize: (1), s.t. (2)-(26) and (28)

## III. Model Reformulation and Solution Method

*A. McCormick Linearization of (25) and (26)*

Both (25) and (26) contain four binary-continuous bi-linear terms. The McCormick relaxation [9] is used to linearize the binary-continuous bi-linear terms without any error.

Here, a general term $\alpha T$ is used to represent the binary-continuous bi-linear terms in (25). Replace it by a new variable, i.e., $Q = \alpha T$, where $Q$ should satisfy

$$\alpha T_{\min} \leq Q \leq \alpha T_{\max} \tag{29}$$

$$T - (1-\alpha)T_{\max} \leq Q \leq T - (1-\alpha)T_{\min}. \tag{30}$$

where $T_{\min}$ and $T_{\max}$ are the lower and upper bounds of $T$, respectively, $\alpha$ is a binary variable, and $Q$ and $T$ are continuous variables.

When $\alpha = 0$, (29) becomes $0 \leq Q \leq 0$, i.e., $Q = 0$. When $\alpha = 1$, (30) becomes $T \leq Q \leq T$, i.e., $Q = T$. Therefore, (29) and (30) ensure that $Q$ is equivalent to $\alpha T$. That is, (25) can be linearized by replacing each binary-continuous bi-linear term by a new variable, $Q$, subject to (29)-(30), which has no error.

Similarly, a general term $\beta p$ is used to represent the binary-continuous bi-linear terms in (26). Replace it by a new variable, i.e., $S = \beta p$, where $S$ should satisfy

$$0 \leq S \leq \beta p_{\max} \tag{31}$$

$$p - (1-\beta)p_{\max} \leq S \leq p \tag{32}$$

where $p_{\max}$ is the upper bound of $p$, $\beta$ is a binary variable, and $S$ and $p$ are continuous variables.

When $\beta = 0$, (31) becomes $0 \leq S \leq 0$, i.e., $S = 0$. When $\beta = 1$, (32) becomes $p \leq S \leq p$, i.e., $S = p$. Therefore, (31) and (32) ensure that $S$ is equivalent to $\beta p$. That is, (26) can be linearized by replacing each binary-continuous bi-linear term by a new variable, $S$, subject to (31)-(32), which has no error.

*B. Piecewise Linearization of Continuous Bi-linear Terms*

In (19)-(24), there are continuous bi-linear terms, i.e., a product of two continuous variables. In this subsection, reformulation and piecewise linearization are used to linearize these bi-linear terms and reduce the complexity of solving the whole model given in Section II. Reference [25] compared different formulations of piecewise linear approximations for non-linear functions, including convex combination, multiple choice, incremental, etc., and concluded that the incremental format consumed the least time for all three cases considered. Therefore, piecewise linearization using an incremental format is used in this paper.

The continuous bi-linear term is represented by a general term, $xy$. First, the bi-linear term is reformulated as the difference of two quadratic terms:

$$xy = (x+y)^2/4 - (x-y)^2/4 \tag{33}$$

The right-hand side of (33) is then piecewise linearized. Let $z^+ = (x+y)/2$ and equally divide the range of $z^+$ into $n_i^+$ segments with each divide point represented by $z_i^+, \forall i = 1,2,\cdots,n_i^+,n_i^+ + 1$. The values of $z_i^+$ and $(z_i^+)^2$ can then be obtained. Using a piecewise linearization method with an incremental format, $(x+y)^2/4$ can be represented by the right-hand side of (34) subject to (35)-(37).

$$(z^+)^2 = (z_1^+)^2 + \sum_{i\in\Omega_{i0}^+}((z_{i+1}^+)^2 - (z_i^+)^2)\varphi_i^+ \quad (34)$$

$$z^+ = (x+y)/2 = z_1^+ + \sum_{i\in\Omega_{i0}^+}(z_{i+1}^+ - z_i^+)\varphi_i^+ \quad (35)$$

$$\varphi_{i+1}^+ \le \zeta_i^+ \le \varphi_i^+,\ \ \zeta_i^+ \in \{0,1\}, \forall i \in \Omega_{i1}^+ \quad (36)$$

$$0 \le \varphi_i^+ \le 1,\ \ \forall i \in \Omega_{i0}^+ \quad (37)$$

where $\varphi_i^+$ is a continuous variable while $\zeta_i^+$ is a binary variable, $\Omega_{i0}^+ = \{1,2,\cdots,n_i^+\}$, and $\Omega_{i1}^+ = \{1,2,\cdots,n_i^+ - 1\}$. Constraints (36) and (37) ensure $\varphi_j^+ = \zeta_j^+ = 1, \forall j < i$ if $\varphi_i^+ > 0$.

Similarly, let $z^- = (x-y)/2$ and equally divide the range of $z^-$ into $n_i^-$ segments with each divide point represented by $z_i^-$, $\forall i = 1,2,\cdots,n_i^-,n_i^- + 1$. The values of $z_i^-$ and $(z_i^-)^2$ can then be obtained. Then $(x-y)^2/4$ can be represented by the right-hand side of (38) subject to (39)-(41).

$$(z^-)^2 = (z_1^-)^2 + \sum_{i\in\Omega_{i0}^-}((z_{i+1}^-)^2 - (z_i^-)^2)\varphi_i^- \quad (38)$$

$$z^- = (x-y)/2 = z_1^- + \sum_{i\in\Omega_{i0}^-}(z_{i+1}^- - z_i^-)\varphi_i^- \quad (39)$$

$$\varphi_{i+1}^- \le \zeta_i^- \le \varphi_i^-,\ \ \zeta_i^- \in \{0,1\}, \forall i \in \Omega_{i1}^- \quad (40)$$

$$0 \le \varphi_i^- \le 1,\ \ \forall i \in \Omega_{i0}^- \quad (41)$$

where $\varphi_i^-$ is a continuous variable while $\zeta_i^-$ is a binary variable, $\Omega_{i0}^- = \{1,2,\cdots,n_i^-\}$, and $\Omega_{i1}^- = \{1,2,\cdots,n_i^- - 1\}$. Constraints (39) and (40) ensure $\varphi_j^- = \zeta_j^- = 1, \forall j < i$ if $\varphi_i^- > 0$.

Now, $xy$ can be expressed by $(z^+)^2 - (z^-)^2$, i.e.,

$$xy = (z_1^+)^2 + \sum_{i\in\Omega_{i0}^+}((z_{i+1}^+)^2 - (z_i^+)^2)\varphi_i^+ - (z_1^-)^2 - \sum_{i\in\Omega_{i0}^-}((z_{i+1}^-)^2 - (z_i^-)^2)\varphi_i^- \quad (42)$$

which is a linear expression.

### *C. Strategies to Reduce the Number of Bi-linear Terms*

Now, the optimization model of UC considering CAES can be formed as

Minimize: (1), s.t. (2)-(26) and (28)

where the continuous bi-linear terms in (19), (20), (21), (22), (23), and (24) are represented by the right-hand side of (42) and the binary-continuous bi-linear terms in (25) and (26) are linearized using the McCormick relaxation described in Section III-A. This is referred to as **Linearized Model I**.

Considering that there are $20n_t$ continuous bi-linear terms, Linearized Model I is difficult to solve for a large-scale power system. Note that piecewise linearization of continuous bi-linear terms introduces extra constraints and binary/continuous variables, which increases the complexity of the model. In this regard, three strategies are proposed to reduce the number of both binary-continuous and continuous bi-linear terms:

1) Instead of using (19)-(24) to represent the temperature and pressure, (19), (21), (23) and (43) are used:

$$T_s^{(t+1)ij} m_s^{(t+1)ij} R = p_s^{(t+1)ij} V_s^i,\ \ \forall t \in \Omega_{T0} \quad (43)$$

2) The bi-linear term, $m_s^{tij} T_s^{tij}$, in (19), (21), and (23) is replaced by $p_s^{tij} V_s^i/R$ according to (43). Then (23) becomes a linear constraint:

$$\frac{a_4}{m_{av0}} e^{-a_4} p_s^{tij} V_s^i/R - T_{s,\text{idl}}^{(t+1)ij} - T_{RW} \frac{a_4}{m_{av0}} e^{-a_4} m_s^{tij} + (c_{10}) T_s^{tij} = T_{RW}(-a_4 e^{-a_4} + e^{-a_4} - 1),\ \ \forall t \in \Omega_{T0} \quad (44)$$

3) Constraints (19) and (21) are merged into a single constraint:

$$-m_s^{tij} T_{s,\text{chdch}}^{(t+1)ij} + p_s^{tij} V_s^i/R + (c_1)\dot{m}_{\text{in}}^{(t+1)ij} T_s^{tij} + (c_2) m_s^{tij} \dot{m}_{\text{in}}^{(t+1)ij} + (c_3)\dot{m}_{\text{in}}^{(t+1)ij} + c_8 T_s^{tij} \dot{m}_{\text{out}}^{(t+1)ij} - \Delta t c_4 T_s^{tij} = -c_4 T_{RW} \Delta t,\ \ \forall t \in \Omega_{T0} \quad (45)$$

where $T_{s,\text{chdch}}^{(t+1)ij}$ is used to represent the temperature at time $(t+1)$ if the $t$th period is either a charging or discharging process.

If the $t$th period is a charging process, then the 6th term in (45) is zero as $\dot{m}_{\text{out}}^{(t+1)ij} = 0$, $T_{s,\text{chdch}}^{(t+1)ij}$ is equivalent to $T_{s,\text{ch}}^{(t+1)ij}$, and (45) is equivalent to (19). If the $t$th period is a discharging process, then the 3rd-5th terms in (45) are zero as $\dot{m}_{\text{in}}^{(t+1)ij} = 0$, $T_{s,\text{chdch}}^{(t+1)ij}$ is equivalent to $T_{s,dch}^{(t+1)ij}$, and (45) is equivalent to (21). Using (45) instead of (19) and (21) results in one less continuous bi-linear term.

Now, (25) becomes

$$T_s^{(t+1)ij} = \left(\alpha^{(t+1)ij} + \beta^{(t+1)ij}\right) T_{s,\text{chdch}}^{(t+1)ij} + \left(1 - \alpha^{(t+1)ij} - \beta^{(t+1)ij}\right) T_{s,\text{idl}}^{(t+1)ij}, \forall t \in \Omega_{T0} \quad (46)$$

which can be linearized in a similar way to (29)-(30).

Now, the improved model (referred to as **Linearized Model II**) can be expressed as

Minimize: (1), s.t. (2)-(18), (28), (43)-(46)

where (46) is linearized in a similar way to (29)-(30) and the continuous bi-linear terms in (43) and (45) are represented by the right-hand side of (42).

Linearized Model I linearizes $20n_t$ continuous bi-linear terms while Linearized Model II linearizes $5n_t$ terms, i.e., the number of continuous bi-linear terms is reduced by a factor of four. Moreover, Linearized Model II avoids (26). The advantage of Linearized Model II over Linearized Model I is significant and will be shown in Section IV-D.

### *D. Constant-Temperature Model*

In the literature, the temperature of the air in the cavern is assumed to be constant [15]-[18]. That is, the cavern model can be modeled as

$$p_s^{(t+1)ij} = m_s^{(t+1)ij} R T_{con}^i / V_s^i,\ \ \forall t \in \Omega_{T0} \quad (47)$$

where $T_{con}^i$ represents the temperature of the air in the cavern.

The corresponding **optimization model** of UC considering CAES using a constant air temperature model can be formed as

Minimize: (1), s.t. (2)-(18), (28), (47)

### *E. Method to Obtain an Initial Solution*

Solving Linearized Models I and II for 24 hours is time consuming. To shorten the solution time, a method to obtain an initial solution for both models is proposed; a flowchart of the method is given in Fig. 1.

Step 1 is to solve the constant-temperature cavern model given in Section III-D and obtain a solution, named solution 1. Step 2 is to solve the CAES model (i.e., (10)-(28)) where the charging and discharging power of CAES are fixed to be the same as solution 1. The maximum and minimum air pressures obtained in step 2 are denoted as $p_{\max 1}$ and $p_{\min 1}$, respectively. If both $p_{\max 1}$ and $p_{\min 1}$ are within the optimal operating range as given in (12) and described in box 3 of Fig. 1, then go to step 5; otherwise go to step 4. In step 4, decrease and increase the maximum and minimum pressures at the two ends of (12), respectively, according to box 4 of Fig. 1. In step 5, Linearized Model I or II is solved by setting the charging/discharging power equal to solution 1; the solution obtained in this step is denoted as solution 2. In step 6, solution 2 is used as an initial solution to solve Linearized Model I or II without fixing the charging/discharging power. The solution obtained in step 6 is the final solution of Linearized Model I or II.

Note that when the charging and discharging power are fixed, $m_s^{tij}$, $\dot{m}_{\text{in}}^{(t+1)ij}$, and $\dot{m}_{\text{out}}^{(t+1)ij}$ can be directly calculated from (28), (10), and (11), respectively, i.e., they become parameters instead of variables. Therefore, the bi-linear constraints (19)-(24), (43), and (45)-(46) become linear constraints and can be solved very fast, i.e., steps 2 and 5 consume very little time.

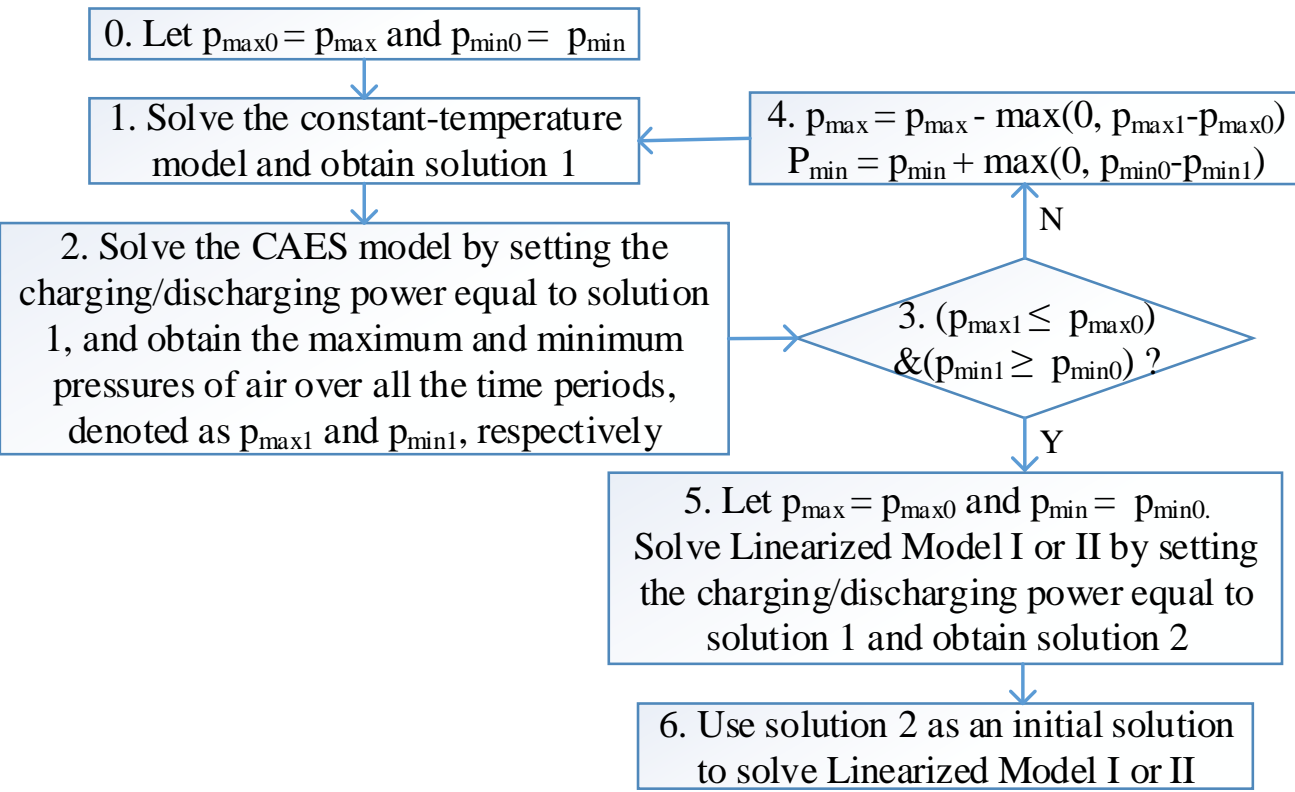

Fig. 1. Flowchart of the solution method using an initial solution.

## IV. SIMULATION

### A. Test System

To verify the effectiveness of the proposed model and solution method, Linearized Models I and II, the constant-temperature model, and the UC model without CAES are solved separately on a modified RTS-79 system with 33 conventional generators [15]. Three same-capacity wind farms are added to the system and located at Buses 1, 4, and 6, respectively. The maximum load demand is set to 3100 MW and the maximum wind penetration is set to 35%. The load and wind profiles are given in Fig. 2. The wind profile comes from the real output of a wind farm in Saskatchewan, Canada. The wind power for scenarios 1 and 3 is set to 0.8 and 1.2 times that of scenario 2, respectively. All of the models are solved using MATLAB® on a Lenovo® ThinkStation with two Intel Xeon E5-2650 V4 processors. Both the charging and discharging costs are set to 3 $/MWh, the wind shedding cost is set to 100 $/MWh, and the reserve cost is set to 3 $/MWh. All other data used can be obtained from the RTS-79 system [15] and MatPower [22]. The parameter of the CAES plant comes from the Huntorf CAES plant as described in the first paper of this two-part series and the optimal operating range of the air pressure in the cavern is 46-66 bar which is used in (12). The linearized model is an MILP problem and is solved using CPLEX. The relative mixed-integer programming (MIP) gap in the CPLEX is set to 0.1%.

The time interval of CAES model is set to 20 minutes which will be further discussed in Section IV-C. The time resolutions for the unit on/off schedule and generation dispatch are one hour and 20 minutes, respectively.

### B. UC with/without CAES

The results obtained from Linearized Model II are given in Figs. 3 and 4a. The total load demand and the total output of all of the conventional generators in the three scenarios are given in Fig. 3a. The total power capacity and the total output of the three wind farms are depicted in Fig. 3b. In scenario 1, all of the wind power can be integrated. Scenario 2 (3) features some (much more) wind shedding. The charging/discharging power of the CAES is given in Fig. 3c. Fig. 3 shows that the CAES discharges in low-wind periods, i.e., periods 32-41 and 50-63, and charges in the other hours. Fig. 4a shows the UC result where each row (column) is associated with a unit (a period of time), and a unit is on (off) if it is filled (blank).

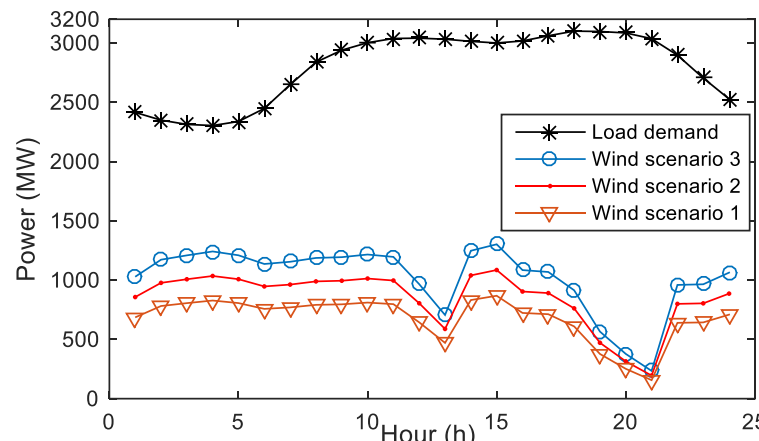

Fig. 2. Load and wind profiles.

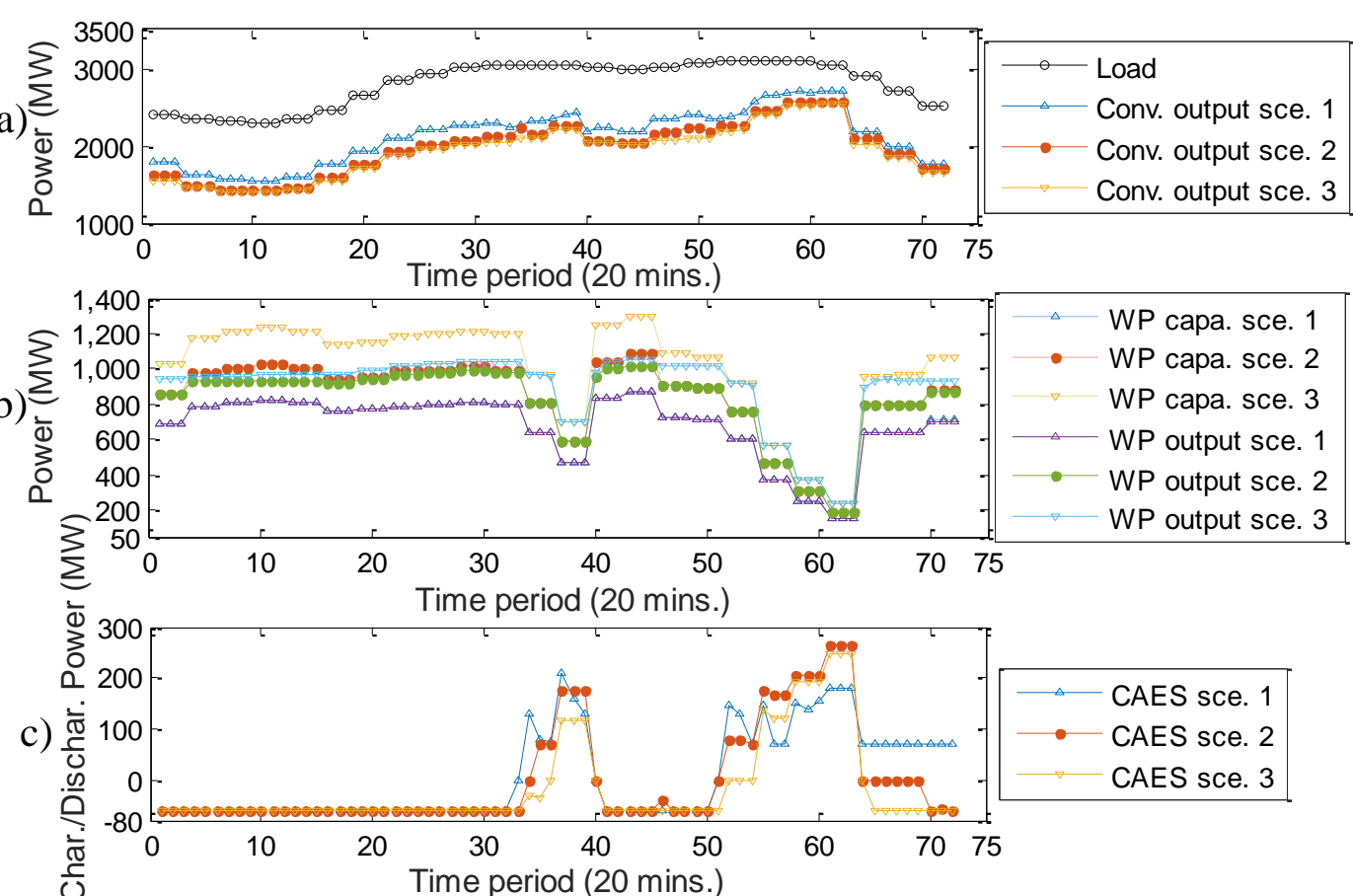

Fig. 3. Results obtained from Linearized Model II: a) load and total output from conventional units, b) total wind power capacity and total wind power output, c) charging/discharging power of CAES.

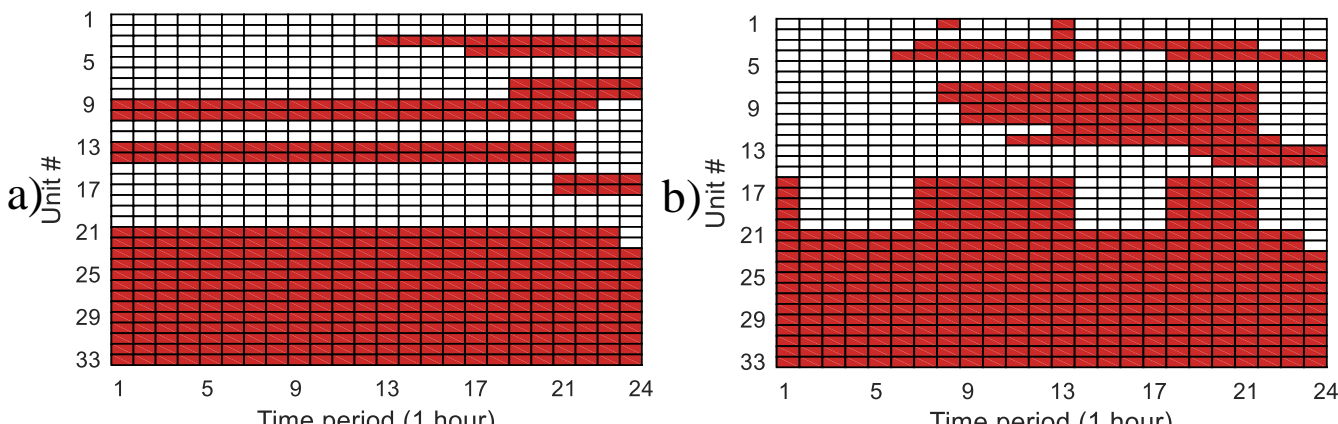

Fig. 4. UC result obtained from a) Linearized Model II and b) the UC model without CAES.

To see the benefit of CAES, the UC without CAES is also solved and the results are shown in Figs. 5 and 4b. Fig. 5 shows that the wind power generation drops in periods 37-39 and 55-63. To satisfy the load in these low-wind periods, more units are turned on to generate more power as shown in Figs. 5a and 4b. Comparing Fig. 3a with Fig. 5a shows that CAES reduces the power output from conventional generators in low-wind periods. Comparing Fig. 3b with Fig. 5b shows that CAES helps to reduce wind shedding, especially in scenarios 2 and 3 that have more wind power. Comparing Figs. 4a with 4b shows that CAES reduces the number of times conventional units are turned on and off.

To investigate the impacts of wind power penetration on the benefits of CAES, the UC problems with and without CAES are solved separately by setting the wind power penetration to 32, 35, and 38%. The results in terms of total cost and wind power shedding are tabulated in Table I, which indicates that CAES can reduce wind power shedding by 392.8, 754.4, and 992.6 MWh and total costs by 2.8, 5.3, and 6.0% for the three different wind power penetrations, respectively. That is, the benefit attributed to CAES increases as the wind power penetration increases.

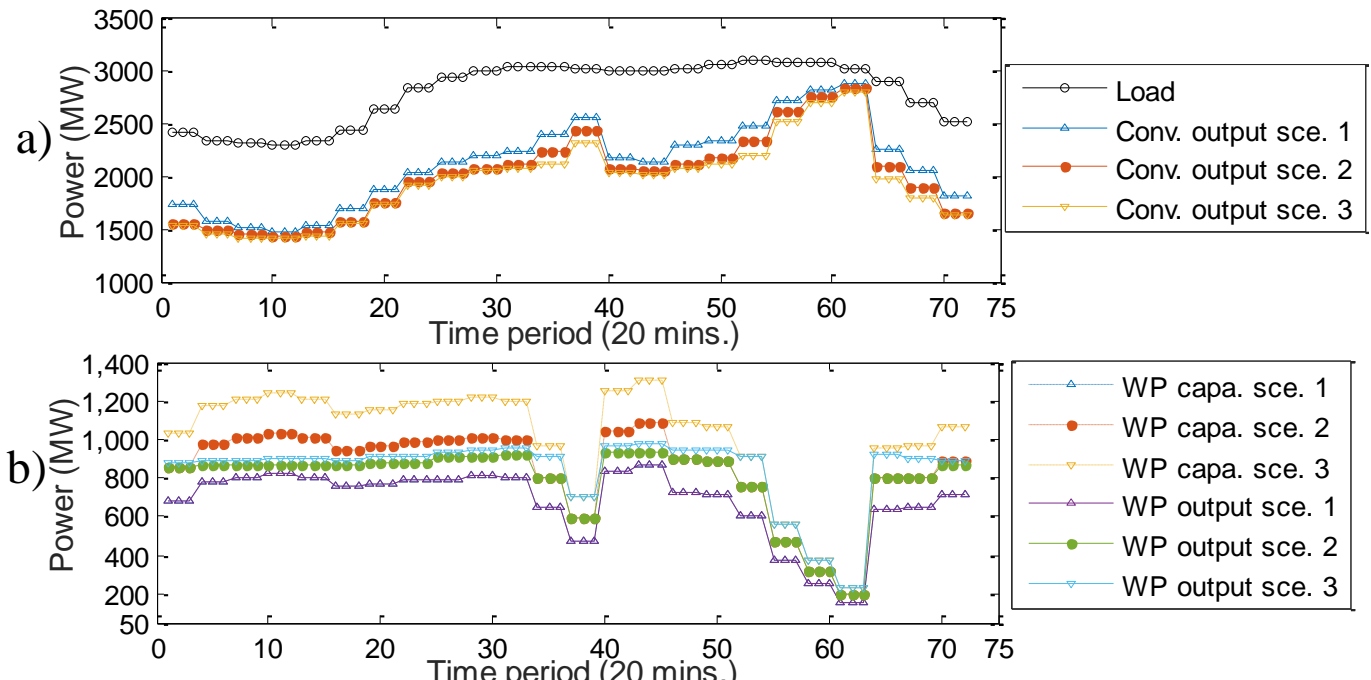

Fig. 5. Results obtained from the UC model without CAES: a) load and total output from conventional units, b) total wind power capacity and total wind power output.

TABLE I
TOTAL COST AND WIND SHEDDING OF SOLUTIONS OBTAINED FROM UC WITH AND WITHOUT CAES UNDER DIFFERENT LEVELS OF WIND POWER PENETRATION.

| Penetration | Linearized Model II | | UC without CAES | |
|---|---|---|---|---|
| | Cost ($) | Wind shed. (MWh) | Cost ($) | Wind shed. (MWh) |
| 32% | 698159 | 313.0 | 717711 | 705.8 |
| 35% | 732886 | 823.9 | 771870 | 1578.3 |
| 38% | 797294 | 1599.0 | 845019 | 2591.6 |

### *C. Comparison Between the Linearized Model II and the Constant-temperature Model*

To show the superiority of the proposed model, the pressure and temperature results obtained from Linearized Model II given in Section III-C (constant-temperature model [16] given in Section III-D) are plotted in Figs. 6a and 6c (Figs. 6b and 6d), respectively. Furthermore, the charging/discharging power obtained from the bi-linear model (constant-temperature model) is used by the accurate model [19] to calculate the pressure and temperature, which are also plotted in Figs. 6a and 6c (Figs. 6b and 6d). That is, the accurate model is used to verify the accuracy of the bi-linear and the constant-temperature models.

Figs. 6a and 6c show that the pressure/temperature obtained from Linearized Model II and the analytical model of CAES [19] are quite close to one another. Note that only scenario 2 is shown in Fig. 6; scenarios 1 and 3 are similar but not shown as the space of the paper is limited. The average relative errors between the pressure (temperature) obtained by the two models are 0.27, 0.28, and 0.28% (0.27, 0.27, and 0.28%) for scenarios 1, 2, and 3, respectively. That is, the bi-linear model is accurate. Therefore, the time interval of the cavern model for CAES can be set to 20 minutes and there is no need to decrease this time interval to further increase accuracy at the expense of a higher computational burden.

Figs. 6b and 6d clearly show that the pressure and temperature obtained from the constant-temperature model are inaccurate. The average relative errors between the pressure (temperature) obtained by the two models are 1.55, 1.49, and 1.39% (1.55, 1.49, and 1.39%) for scenarios 1, 2, and 3, respectively, which are about 5-5.5 times the errors of Linearized Model II. Even worse, the pressure obtained from the accurate model goes below the lower bound (46 bar) of its operating range, i.e., the solution obtained from the constant-temperature model actually allows the cavern of CAES plant to operate outside the optimal pressure region (i.e., 46-66 bar). However, the solution obtained from the bi-linear model ensures the cavern of CAES plant operates within the optimal pressure region. Therefore, it is necessary to use the proposed Linearized Model II to obtain an accurate and feasible solution.

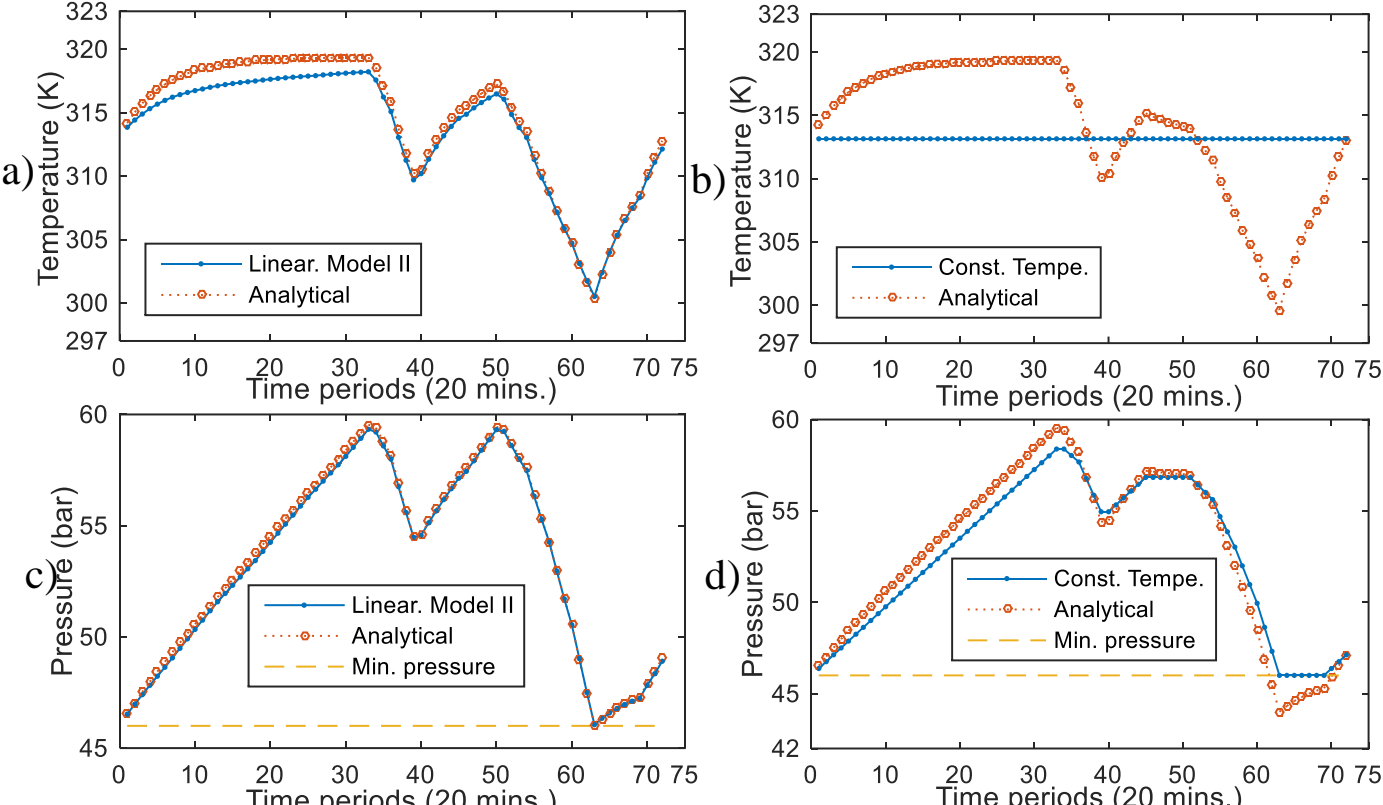

Fig. 6. Temperature result obtained from a) Linearized Model II and the analytical model [19], and b) the constant-temperature model and the analytical model [19]; Pressure result obtained from c) Linearized Model II and the analytical model [19], and d) the constant-temperature model and the analytical model [19].

### *D. Comparison Between Linearized Model II and Linearized Model I*

To show the effectiveness of the proposed Linearized Model II and the method to obtain an initial solution, the time consumed to solve the linearized models directly or using the method to obtain an initial solution as described in Section III-E is tabulated in Table II (where '----' indicates 'it does not converge after running for 7 days'). In Table II, $t_{\mathrm{dir}}$ and $t_{\mathrm{ini}}$ represent the time consumed to solve Linearized Model II directly and to obtain initial solution, respectively; $t_{6,\mathrm{LM-I}}$ and $t_{6,\mathrm{LM-II}}$ represent the time consumed in step 6 to solve Linearized Model I and Linearized Model II, respectively.

Note that the final solution of Linearized Model II given in previous subsections is obtained in step 6, as described in Section III-E. The initial solution does not affect the optimality of the final solution of the linearized whole model as the optimality is determined by the termination condition of the MILP solver, i.e., the relative MIP gap goes below 0.1%.

The 2nd column of Table II shows that solving Linearized Model II directly is fast when the number of hours is small but intractable as the number of hours increases. Note that it is more difficult to solve the model as the number of hours increases. The 3rd and 4th columns of Table II show that the initial solution can be obtained in a relatively short time and that it is a near-optimal solution with an optimality gap of around 3%. The last two columns of Table II show that solving Linearized Model II is much easier than Linearized Model I, especially when the number of hours is large, which indicates the effectiveness of the proposed three strategies. Comparing the 6th and the 2nd columns indicates that the initial solution significantly reduces the solution time. Therefore, the three strategies, linearization, and the initial solution are quite effective and necessary, which helps to solve the whole model effectively by converting it into Linearized Model II.

TABLE II
TIME CONSUMED TO SOLVE THE LINEARIZED MODELS DIRECTLY OR USING AN INITIAL SOLUTION.

| No. of hours | $t_{\text{dir}}$ | $t_{\text{ini}}$ | Gap of initial solu. | $t_{6,\text{LM}-\text{I}}$ | $t_{6,\text{LM}-\text{II}}$ |
|---|---|---|---|---|---|
| 3 | 22 s | 2.9 s | 3.00% | 58 s | 11 s |
| 5 | 37673 s | 5.7 s | 3.10% | 27248 s | 46 s |
| 24 | ---- | 319 s | 2.87% | ---- | 1565 s |

## V. CONCLUSION

A UC model considering CAES has been proposed in this paper using the accurate bi-linear model proposed in the first paper of this two-part series. Simulation results show that the bi-linear cavern model is more accurate and avoids violating the optimal operating range of the air pressure in the cavern compared to the constant-temperature cavern model. Therefore, it is necessary and beneficial to use the accurate bi-linear cavern model.

However, the bi-linear terms complicate the whole model. To address this issue, three strategies have been proposed to reduce the number of bi-linear terms in the whole model. Thereafter, the McCormick relaxation and piecewise linearization are used to linearize the binary-continuous and continuous bi-linear terms, respectively. Moreover, a method to generate an initial solution, based on the solution of the UC with CAES using a constant-temperature cavern model, for the whole model has been proposed.

Simulation results show that the three strategies reduce the complexity of the whole model and, hence, significantly reduce the solution time required to solve the linearized whole model; the initial solution also substantially reduces the solution time; the whole model can be effectively solved by using the three strategies, linearization, and the initial solution. Simulation results also show that integrating CAES in the UC problem reduces wind shedding, total cost, and the number of times conventional generators are turned on and off. The benefit of CAES increases as the penetration of wind power increases.

## APPENDIX

$a_2 = \frac{R^k T_{in}^k}{V_s^k p_{in}^{k-1}}$, $a_3 = \frac{R^{k-1} T_{in}^k}{V_s^{k-1} p_{in}^{k-1}}$, $c_1 = (k-2)\Delta t - 0.5(k-2)\Delta t^2/(c_v m_{av0})$,

$a_4 = \frac{h_c A_c \Delta t}{m_{av0} c_v}$, $c_2 = a_3 \Delta t (k-1)(m_{av0})^{k-2} - (k-2)(m_{av0})^{k-3} \frac{c_4}{2} a_3 \Delta t^2$,

$c_3 = a_3 \Delta t (m_{av0})^{k-1} - (c_2) m_{av0} - (m_{av0})^{k-2} c_4 0.5 a_3 \Delta t^2$, $c_4 = \frac{h_c A_c}{c_v}$,

$c_5 = -0.5 c_4 (k-1) \Delta t^2 R / V_s^i$, $c_7 = 0.5 c_4 T_{RW} \Delta t^2 R / V_s^i + (1-k) a_2 \Delta t m_{av0}^k$,

$c_6 = c_4 T_{RW} \Delta t R / V_s^i$, $c_8 = (c_4 \Delta t^2 / (2 m_{av0}) - \Delta t)(k-1)$,

$c_9 = 0.5 c_4 R \Delta t^2 k / V_s^i$, $c_{10} = e^{-a_4} - a_4 e^{-a_4}$, $c_{11} = (1 - c_{10}) R T_{RW} / V_s^i$.

## REFERENCES

[1] J. Wang, M. Shahidehpour, and Z. Li, "Security-constrained unit commitment with volatile wind power generation", *IEEE Trans. Power Syst.*, vol. 23, no. 3, pp. 1319-1327, 2008.

[2] F. Aminifar, M. Fotuhi-Firuzabad, and M. Shahidehpour, "Unit commitment with probabilistic spinning reserve and interruptible load considerations", *IEEE Trans. Power Syst.*, vol. 24, no. 1, pp. 388-397, 2009.

[3] Y. Wen, C. Guo, H. Pandžić, and D.S. Kirschen, "Enhanced security-constrained unit commitment with emerging utility-scale energy storage", *IEEE Trans. Power Syst.*, vol. 31, no. 1, pp. 652-662, 2016.

[4] H. Pandžić, Y. Dvorkin, T. Qiu, Y. Wang, and D.S. Kirschen, "Toward cost-efficient and reliable unit commitment under uncertainty", *IEEE Trans. Power Syst.*, vol. 31, no. 2, pp. 970-982, 2016.

[5] N.P. Padhy, "Unit commitment-a bibliographical survey", *IEEE Trans. Power Syst.*, vol. 19, no. 2, pp. 1196-1205, 2004.

[6] Q.P. Zheng, J. Wang, and A.L. Liu, "Stochastic optimization for unit commitment - a review", *IEEE Trans. Power Syst.*, vol. 30, no. 4, pp. 1913-1924, 2015.

[7] R. Jiang, J. Wang, and Y. Guan, "Robust unit commitment with wind power and pumped storage hydro", *IEEE Trans. Power Syst.*, vol. 27, no. 2, pp. 800-810, 2012.

[8] K. Bruninx, Y. Dvorkin, E. Delarue, H. Pandžić, W.D. Haeseleer, and D.S. Kirschen, "Coupling pumped hydro energy storage with unit commitment", *IEEE Trans. Sustainable Energy*, vol. 7, no. 2, pp. 786-796, 2016.

[9] Y. Wen, W. Li, G. Huang, and X. Liu, "Frequency dynamics constrained unit commitment with battery energy storage", *IEEE Trans. Power Syst.*, vol. 31, no. 6, pp. 5115-5125, 2016.

[10] B. Cleary, A. Duffy, A. O'Connor, M. Conlon, and V. Fthenakis, "Assessing the economic benefits of compressed air energy storage for mitigating wind curtailment", *IEEE Trans. Sustainable Energy*, vol. 6, no. 3, pp. 1021-1028, 2015.

[11] B.M. Enis, P. Lieberman, and I. Rubin, "Operation of hybrid wind-turbine compressed-air system for connection to electric grid networks and cogeneration", *Wind Engineering*, vol. 27, no. 6, pp. 449-459, 2003.

[12] H. Khani, M.R.D. Zadeh, and A.H. Hajimiragha, "Transmission congestion relief using privately owned large-scale energy storage systems in a competitive electricity market", *IEEE Trans. Power Syst.*, vol. 31, no. 2, pp. 1449-1458, 2016.

[13] S. Shafiee, H. Zareipour, A.M. Knight, N. Amjady, and B. Mohammadi-Ivatloo, "Risk-constrained bidding and offering strategy for a merchant compressed air energy storage plant", *IEEE Trans. Power Syst.*, vol. 32, no. 2, pp. 946-957, 2017.

[14] S. Shafiee, H. Zareipour, and A. Knight, "Considering thermodynamic characteristics of a CAES facility in self-scheduling in energy and reserve markets", *IEEE Trans. Smart Grid*, Early access.

[15] D. Pozo, J. Contreras, and E.E. Sauma, "Unit commitment with ideal and generic energy storage units", *IEEE Trans. Power Syst.*, vol. 29, no. 6, pp. 2974-2984, 2014.

[16] H. Daneshi, and A.K. Srivastava, "Security-constrained unit commitment with wind generation and compressed air energy storage", *IET Gener. Transm. Distrib.*, vol. 6, no. 2, pp. 167-175, 2012.

[17] H. Bitaraf, H. Zhong, and S. Rahman, "Managing large scale energy storage units to mitigate high wind penetration challenges," in *IEEE PES General Meeting*, Denver, CO, USA, pp. 1-5, Jul. 2015.

[18] J. Zhang, K.J. Li, M. Wang, W.J. Lee, and H. Gao, "A bi-level program for the planning of an islanded microgrid including CAES," in *IEEE Ind. Appl. Society Annual Meeting*, Addison, TX, USA, pp. 1-8, Oct. 2015.

[19] C. Xia, Y. Zhou, S. Zhou, P. Zhang, and F. Wang, "A simplified and unified analytical solution for temperature and pressure variations in compressed air energy storage caverns", *Renewable Energy*, vol. 74, pp. 718-726, 2015.

[20] Geißler, B., Martin, A., Morsi, A., Schewe, L. Using piecewise linear functions for solving MINLPs. In: Mixed Integer Nonlinear Programming. Springer, pp. 287–314, 2012.

[21] E. Klotz, and A.M. Newman, "Practical guidelines for solving difficult mixed integer linear programs", *Surveys in Operations Res. and Management Science*, vol. 18, no. 1, pp. 18-32, 2013.

[22] C.E. Murillo-Sánchez, R.D. Zimmerman, C.L. Anderson, and R.J. Thomas, "Secure planning and operations of systems with stochastic sources, energy storage, and active demand", *IEEE Trans. Smart Grid*, vol. 4, no. 4, pp. 2220-2229, 2013.

[23] C.Y. Chung, H. Yu, and K.P. Wong, "An advanced quantum-inspired evolutionary algorithm for unit commitment", *IEEE Trans. Power Syst.*, vol. 26, no. 2, pp. 847-854, 2011.

[24] T. Das, V. Krishnan, Y. Gu, and J. D. McCalley, "Compressed air energy storage: state space modeling and performance analysis," in *IEEE PES General Meeting,* Detroit, USA, pp.1-8, Jul. 2011.

[25] C.M. Correa-Posadaa, and P. Sanchez-Martin, "Gas network optimization: a comparison of piecewise linear models," *Chemical Engineering Science*. pp.1-24, Jun. 2014.